\documentclass[apj]{emulateapj}
\usepackage{mathptmx}

\newcommand  \kms      {\ifmmode {\rm km\,s}^{-1} \else km\,s$^{-1}$\fi}
\newcommand  \cc       {\hbox{cm$^{-3}$}}
\newcommand  \cmii     {\hbox{cm$^{-2}$}}
\newcommand  \ergs     {\ifmmode {\rm ergs\,s}^{-1} \else ergs s$^{-1}$\fi}
\newcommand  \ergcms   {\ifmmode {\rm ergs\,cm}^{-2}\,{\rm s}^{-1}
                        \else ergs\,cm$^{-2}$\,s$^{-1}$\fi}
\newcommand  \ergcmsA {\ifmmode{\rm ergs\,cm}^{-2}\,{\rm s}^{-1}\,{\rm\AA}^{-1}
                        \else ergs\,cm$^{-2}$\,s$^{-1}$\,\AA$^{-1}$\fi}
\newcommand \ergcmsHz {\ifmmode{\rm ergs\,cm}^{-2}\,{\rm s}^{-1}\,{\rm Hz}^{-1}
                        \else ergs\,cm$^{-2}$\,s$^{-1}$\,Hz$^{-1}$\fi}
\newcommand  \phcms    {\ifmmode {\rm ph\,cm}^{-2}\,{\rm s}^{-1}
                        \else ,ph\,cm$^{-2}$\,s$^{-1}$\fi}
\newcommand  \phcmsA   {\ifmmode {\rm ph\,cm}^{-2}\,{\rm s}^{-1}\,{\rm\AA}^{-1}
                        \else ph\,cm$^{-2}$\,s$^{-1}$\,\AA$^{-1}$\fi}
\def\micron{\ifmmode \mu{\rm m} \else $\mu$m\fi}
\def\kms{\ifmmode {\rm km\,s}^{-1} \else km\,s$^{-1}$\fi}
\def\Hubble{\ifmmode {\rm km\,s}^{-1}\,{\rm Mpc}^{-1}
        \else km\,s$^{-1}$\,Mpc$^{-1}$\fi}
\def\ergsec{\ifmmode {\rm ergs\;s}^{-1} \else ergs s$^{-1}$\fi}
\def\ergscm{\ifmmode {\rm ergs\,s}^{-1}\,{\rm cm}^{-2}
          \else ergs\,s$^{-1}$\,cm$^{-2}$\fi}
\def\ergscmA{\ifmmode {\rm ergs\,s}^{-1}\,{\rm cm}^{-2}\,{\rm \AA}^{-1}
          \else ergs\,s$^{-1}$\,cm$^{-2}$\,\AA$^{-1}$\fi}
\def\ergscmHz{\ifmmode {\rm ergs\,s}^{-1}\,{\rm cm}^{-2}\,{\rm Hz}^{-1}
          \else ergs\,s$^{-1}$\,cm$^{-2}$\,Hz$^{-1}$\fi}
\def\Msun{\ifmmode M_{\odot} \else $M_{\odot}$\fi}
\def\msun{\ifmmode M_{\odot} \else $M_{\odot}$\fi}
\def\Lsun{\ifmmode L_{\odot} \else $L_{\odot}$\fi}
\def\lsun{\ifmmode L_{\odot} \else $L_{\odot}$\fi}
\def\qo{\ifmmode q_{0} \else $q_{0}$\fi}
\def\Ho{\ifmmode H_{0} \else $H_{0}$\fi}
\def\ho{\ifmmode h_{0} \else $h_{0}$\fi}
\def\qo{\ifmmode q_{0} \else $q_{0}$\fi}
\def\ao{\ifmmode a_{0} \else $a_{0}$\fi}
\def\to{\ifmmode t_{0} \else $t_{0}$\fi}
\def\Halpha{\ifmmode {\rm H}\alpha \else H$\alpha$\fi}
\def\Hbeta{\ifmmode {\rm H}\beta \else H$\beta$\fi}
\def\hb{\ifmmode {\rm H}\beta \else H$\beta$\fi}
\def\Hgamma{\ifmmode {\rm H}\gamma \else H$\gamma$\fi}
\def\Hdelta{\ifmmode {\rm H}\delta \else H$\delta$\fi}
\def\Lya{\ifmmode {\rm Ly}\alpha \else Ly$\alpha$\fi}
\def\Lyb{\ifmmode {\rm Ly}\beta \else Ly$\beta$\fi}
\def\hi{\ifmmode \mbox{{\rm H}\,{\sc i}} \else H\,{\sc i}\fi}

\def\ciii{\ifmmode {\rm C}\,{\sc iii} \else C\,{\sc iii}\fi}

\def\oviii{O\,{\sc viii}}

\def\neix{Ne\,{\sc ix}}
\def\nex{Ne\,{\sc x}}

\def\mgXI{Mg\,{\sc xi}}
\def\mgXII{Mg\,{\sc xii}}
\def\mgxii{Mg\,{\sc xii}}

\def\siXIII{Si\,{\sc xiii}}
\def\siXIV{Si\,{\sc xiv}}
\def\sXV{S\,{\sc xv}}
\def\sXVI{S\,{\sc xvi}}

\def\sXV{S\,{\sc xv}}
\def\sXVI{S\,{\sc xvi}}

\def\feXVII{Fe\,{\sc xvii}}

\def\feXXV{Fe\,{\sc xxv}}
\def\feXXVI{Fe\,{\sc xxvi}}

\def\o5007{[O\,{\sc iii}]\,$\lambda5007$}
\def  \kms         {\hbox{km s$^{-1}$}}          
\def  \ergs        {\hbox{erg s$^{-1}$}}              
\def  \cc          {\hbox{cm$^{-3}$}}

\def  \cmii        {\hbox{cm$^{-2}$}}

\def  \etal        {{\rm et al.}}
\def  \La          {\ifmmode {\rm Ly}\alpha \else Ly$\alpha$\fi}
\def  \Ka          {\ifmmode {\rm K}\alpha \else K$\alpha$\fi}
\def  \Kb          {\ifmmode {\rm K}\beta \else K$\beta$\fi}
\def  \Lb          {\ifmmode {\rm L}\beta \else L$\beta$\fi}
\def  \Ha          {\ifmmode {\rm H}\alpha \else H$\alpha$\fi}
\def  \Hb          {\ifmmode {\rm H}\beta \else H$\beta$\fi}
\def  \Pa          {\ifmmode {\rm P}\alpha \else P$\alpha$\fi}
\def  \CIIIb       {\ifmmode {\rm C}\,{\sc iii]}\,\lambda1909
                     \else C\,{\sc iii]}\,$\lambda1909$\fi}
\def  \CIV         {\ifmmode {\rm C}\,{\sc iv}\,\lambda1549
                     \else C\,{\sc iv}\,$\lambda1549$\fi}
\def  \MgII         {\ifmmode {\rm Mg}\,{\sc ii}\,\lambda2798
                     \else Mg\,{\sc ii}\,$\lambda2798$\fi}
\def  \OVI         {\ifmmode {\rm O}\,{\sc vi}\,\lambda1035
x
                     \else O\,{\sc vi}\,$\lambda1035$\fi}
\def \chandra  {{\it Chandra}}
\def \xmm      {{\it XMM-Newton}}

\journalinfo{The Astrophysical Journal, 
6??: 1--10, 2005 Month, astro-ph/0505016}
\slugcomment{Received 2005 January 24; accepted 2005 April 23}

\shorttitle{{\it XMM-NEWTON} SPECTROSCOPY OF NGC\,6240}
\shortauthors{NETZER ET AL.}

\begin{document}

\title{{\it XMM-NEWTON} Spectroscopy of the Starburst Dominated Ultra Luminous  \\ Infrared Galaxy NGC\,6240}

\author{
Hagai Netzer,\altaffilmark{1}
Doron Lemze,\altaffilmark{1}
Shai Kaspi,\altaffilmark{1,2}
I.M.~George,\altaffilmark{3,4}
T.J.~Turner,\altaffilmark{3,4} \\
D.~Lutz,\altaffilmark{5}
T.~Boller,\altaffilmark{5}
and Doron Chelouche \altaffilmark{6,7}
}

\altaffiltext{1}
                {School of Physics and Astronomy and the Wise
                Observatory, The Raymond and Beverly Sackler Faculty of
                Exact Sciences, Tel-Aviv University, Tel-Aviv 69978,
                Israel; netzer@wise.tau.ac.il}

\altaffiltext{2}
                {Physics Department, The Technion, Haifa 3200, Israel}

\altaffiltext{3}
                {Physics Department, University of Maryland Baltimore County, Baltimore MD 21250, USA}

\altaffiltext{4}
                {Exploration of the Universe Division, Code 660, NASA/Goddard Space Flight Center, Greenbelt, MD 20771.}
\altaffiltext{5} {Max-Planck-Institut f\"ur extraterrestrische Physik, Postfach 1312, 85741 Garching, Germany}
\altaffiltext{6} {School of Natural Sciences, Institute for Advanced Study,  Einstein Dr.  Princeton,  NJ 08540, USA}
\altaffiltext{7} {Chandra Fellow}

\begin{abstract}

We present new \xmm\ observation of the Ultra Luminous Infrared Galaxy (ULIRG)
NGC\,6240. We analyze the reflecting grating spectrometer (RGS) data, and data from the other instruments,
and find a starburst dominated 0.5--3 keV spectrum with global properties resembling those
observed in M82 but with a  much higher luminosity. We show that the starburst region
can be divided into an outer zone, beyond a radius of
about 2.1 kpc, with a gas temperature of about $10^7$ K and a central region
with temperatures in the range (2--6)$\times 10^7$~K.
The gas in the outer region emits most of the
observed \oviii\ $L_{\alpha}$ line and the gas in the inner region the emission lines of higher ionization
 ions, including a strong \feXXV\ line. We also identify a small inner part, very
close to the active nuclei, with typical Seyfert 2 properties including a large amount of
 photoionized gas producing  a strong Fe \Ka\ 6.4 keV line.
The combined abundance, temperature and emission measure analysis indicates super solar
Ne/O, Mg/O, Si/O, S/O and possibly also Fe/O.
The analysis suggests densities in the
range of (0.07--0.28)$\epsilon^{-1/2}$ \cc\ and a total thermal gas mass of $\sim 4 \times 10^8 \epsilon^{1/2}$ \msun, where
$\epsilon$ is the volume filling factor.
 We used a simple model to argue that a massive starburst with an age of
 $\simeq 2 \times 10^7$ years can explain most of the observed properties of the source.
NGC\,6240 is perhaps the clearest case of an X-ray bright luminous AGN, in a merger, whose soft X-ray
spectrum is dominated by a powerful starburst.

\end{abstract}


\section{Introduction}

Ultra-luminous infrared  galaxies (ULIRGs) have most of their
luminosity emerging in the infrared with
L(8-1000~$\mu m)>  3 \times 10^{45}$ \ergs\ (Sanders \& Mirabel 1996).
Many ULIRGs occur in interacting systems and it is thought
that galaxy mergers produce concentrations of gas which  can fuel
a giant black hole (BH)  and/or result in prodigious starburst activity.
ULIRGs are fundamental for the understanding of the star formation history in
the universe, black hole merging and growth and as a trigger of activity in
galactic nuclei.
Yet, the relationship between these systems,   ``normal''
active galactic nuclei (AGN) and
starburst galaxies is still unclear (e.g. Sanders \etal\ 1988; Genzel
et al., 1998; Levenson, Weaver and Heckman 2001a).
 Of special interest are those ULIRGs showing
a clear AGN signature in the X-ray but only very weak AGN-type emission in the IR.
 This may indicate almost complete obscuration of the central AGN
with important implications to the X-ray background and AGN evolution.

NGC~6240 (z=0.0245; hereafter we assume a luminosity distance of 106 Mpc and angular scale such that 1\arcsec\
corresponds to 490 pc at the source) shows two compact optical-IR-radio sources (Tacconi \etal\ 1999) and
two point-like X-ray sources (Komossa et al. 2003, hereafter K03). Its
IR luminosity, using the Sanders and Mirabel (1996) definition and the IRAS fluxes, is
L(8-1000~$\mu m) \simeq  7 \times 10^{11}$ \lsun\ which we assume to represent about 90\% of the bolometric luminosity.
While being a bit
below the ``standard'' ULIRG luminosity of $10^{12}$ \Lsun, the source  has been considered by many
to be a member of this group.
Tecza et al. (2000) show IR imaging spectroscopy of the source and discuss the starburst
in the inner part. More recent IR data are given by Bogdanovic et al. (2003) and by the recent
 high spatial resolution near IR imaging of Max et al. (2004) that shows several
point sources, possibly young star clusters, in the innermost region.
ASCA (Turner \etal\ 1997; Netzer \etal\ 1998), BeppoSax (Vignati \etal\ 1999)
and \chandra\ (K03 and Ptak et al. 2003) observations show the presence of an absorbed
Seyfert 2 type continuum with an absorbing line-of-sight column of
$N_H \sim 1.5 \times 10^{24}~ {\rm cm}^{-2}$ and several large equivalent width (EW) lines. The
absorption corrected X-ray luminosity is
$L_{2-10\,keV} \sim  4 \times 10^{43}\, {\rm erg~s^{-1}}$ (obtained from our own analysis, depending somewhat
on the assumed absorbing column, see below).
Extrapolation of the unobscured X-ray continuum to lower
energies, assuming a ``typical'' AGN spectral energy distribution (SED), suggests that some
50\% of the bolometric luminosity is due to this component, yet there are
only weak high excitation AGN-type IR lines (Genzel \etal\ 1998, Lutz \etal, 2003).

\begin{deluxetable*}{ccccccccc}
\tablecolumns{9}
\tabletypesize{\footnotesize}
\tablewidth{400pt}
\tablecaption{{\it XMM-Newton} Observation Log for NGC\,6240
\label{obslog}}
\tablehead{
\colhead{Obs. ID} &
\colhead{Revolution} &
\colhead{Start time} &
\colhead{Duration} &
\multicolumn{5}{c}{Good time interval (GTI) [ksec]} \\
\colhead{} &
\colhead{} &
\colhead{} &
\colhead{[ksec]} &
\colhead{pn} &
\colhead{MOS1} &
\colhead{MOS2} &
\colhead{RGS1}  &
\colhead{RGS2}   }
\startdata
0101640101 & 0144 & 2000 Sep 22 01:38:46 & 30.1 & 12.3   & 23.6& 23.6&\nodata&\nodata\\
0101640601 & 0413 & 2002 Mar 12 21:37:24 & 19.0 &  7.3   & 15.5  & 16.2  & 10.1 & 9.8\\
0147420201 & 0597 & 2003 Mar 14 18:06:14 & 31.6 &  5.1   & 10.5  & 10.5  & 14.3 &14.3\\
0147420301\tablenotemark{a} & 0599 & 2003 Mar 18 21:01:56 & 28.1 &\nodata &\nodata&\nodata&  2.9 & 2.9\\
0147420401 & 0673 & 2003 Aug 13 10:29:32 & 14.1 &  7.7   & 11.8  & 11.8  &  8.5 & 8.5\\
0147420501 & 0677 & 2003 Aug 21 10:25:09 & 31.2 &  3.6   &\nodata&\nodata& 12.0 &12.0\\
0147420601 & 0681 & 2003 Aug 29 11:25:15 & 9.2  &  1.4   &\nodata&\nodata&  2.9 & 2.9\\
\multicolumn{3}{r}{Total GTI (ks)}     &163.3 & 37.3   &  61.4 &  62.0 & 50.6 &50.3
\enddata
\tablenotetext{a}{EPIC-pn data suffered from telemetry problems, and could not be recovered.}
\end{deluxetable*}

X-ray observations of several well studied type-II AGN
 (e.g. NGC\,1068, Mrk\,3, the Circinus galaxy, see e.g. Kinkhabwala et al. 2002)
 show emission lines from gas which is photoionized by the central non-thermal source.
The starburst contribution to the X-ray emission in those sources is negligible.
Different conclusions about the origin of the X-ray gas in many sources that show combined
Seyfert-starburst signs were reached by Levenson et al. (2001a; 2001b) in their extensive
study of such sources. Unfortunately, all the Levenson et al. spectroscopic analysis is based
on low resolution CCD spectra, mostly by ASCA and ROSAT, which can be very problematic. The clearest
example is NGC\,1068 where Kinkhabwala et al. (2002) followed suggestions by
Marshall et al. (1993) and Netzer and Turner (1997) and  showed, very convincingly,
that the so-called ``thermal X-ray gas''  is in fact photoionized gas in the source.
There are also several well documented cases where abundances as small as 0.1 solar
have been derived from  highly significant fits to CCD spectra  (e.g. Dahlem et al., 1998) while later
works showed that the reason was the lack of spectral resolution and not the unusual abundances.
Apparently, the improved spectral capability provided by \xmm\ and \chandra\ can completely change the
conclusions regarding the origin of the emission line gas in such sources.

 Based on its hard X-ray spectrum, NGC\,6240 can definitely be classified as a very luminous type-II AGN.
 Yet Boller et al. (2003) and Ptak et al. (2003) argued for a more
complex model involving a combination of an
 AGN and a starburst.
Given the need for a high resolution X-ray spectroscopy, we have obtained
 data that enabled us to resolve the AGN-starburst confusion in this source.
\S2\ below discusses new \xmm\ observations of NGC\,6240 and \S3 gives  detailed spectroscopic
analysis and  physical interpretation of the spatially resolved emission in this source.
Finally, in \S4, we discuss implications to the local starburst as well as
to the general question of starburst activity in giant mergers.

\section{Observations and Reduction}

Table~\ref{obslog} gives a summary of the \xmm\ data used in this
paper. It includes all the relevant observations including those published by Boller et
al. (2003, first entry in the table). We reduced
all data using the Science Analysis Software (SAS v5.3.0)
in the standard processing chains as described in the data
analysis threads and the ABC Guide to {\it XMM-Newton} Data
Analysis.\footnote{http://heasarc.gsfc.nasa.gov/docs/xmm/abc}

Unfortunately, all the 2003 observations had  very restricted visibility
windows and had to be split into several short pointings. Also, the observations
were taken at times when the spacecraft was close to the Earth's radiation belts
 resulting in high background count rates.
Several of the observations were thus terminated before their completion.
Table~\ref{obslog} lists the total exposure time of each {\it XMM-Newton} observation and the good time interval
(GTI) for each of the instruments.

The EPIC-pn (hereafter pn) and the two EPIC-MOS (hereafter MOS) detectors were operated during our
2003 observations in the large window mode with the thin
filter. The pn and MOS data were processed using SAS in the standard
way. Source events were extracted from circular regions of
radius 30$\arcsec$ and background events from an annulus with a
radius of 30$\arcsec$--87$\arcsec$. Epochs of high background
event were excluded.
The known PSF properties suggest that some 10\%
of the flux may be emitted outside of the chosen extraction window. We have therefore re-extracted
our longest GTI exposure with twice the above radius and a background region far removed from the source.
The comparison with the results of the smaller extraction region
show an insignificant difference thus the chosen extraction window
has little influence on our final results.

Six pn and four MOS observations, with a total GTI of 37 and 61 ks, respectively, resulted in useful data.
After reducing a spectrum from each observation and verifying
there is no time variability (see below) we combined all spectra
from each detector into one spectrum. This was done using the FTOOL
task MATHPHA. We also combined the matching rmf and arf files using
the tools ADDRMF and ADDARF (weighting individual exposures by their
respective integration times). We then binned the
combined spectrum of each detector to have at least 25 counts per bin.
All data presented here are corrected for galactic absorption with a column density of
$5.9 \times 10^{20}$ \cmii\ (Dickey \& Lockman 1990)

The RGS1 and RGS2 were operated in the standard
spectroscopy mode and produced 50 ks of GTI. Background subtraction was
performed with the SAS using regions adjacent to those containing
the source in the spatial and spectral domains (using the defaults given by the SAS rgsproc
tool). The spectra were
extracted into uniform bins of $\sim$\,0.04 \AA\ (roughly half
the RGS resolution and four times the default bin width) in order to
increase the signal-to-noise (S/N) ratio. Flux calibration was obtained
by dividing the counts by the exposure time and the effective area
at each wavelength. Each flux-calibrated spectrum was also corrected
for Galactic absorption and the two RGS spectra of each observation
were combined into a weighted mean. At wavelengths where the
RGS2 bins did not match exactly the wavelength of the RGS1 bins, we
interpolated the RGS2 data to enable the averaging.  After verifying
that all spectra from different observations are consistent with
each other, we combined the spectra from the six observations listed
in Table~\ref{obslog} into an error-weighted mean which is
presented and analyzed below. At $\lambda > 20$~\AA\ the RGS spectrum is so noisy
that no meaningful line intensity can be measured and even upper limits on emission
line fluxes are highly questionable. Those long wavelength lines definitely require
longer integration.

The spatial resolution of the
\xmm\ instruments was discussed, in great detail, by Boller et al. (2003). These
authors show that some 90\% of the flux is encircled within a radius of about 30\arcsec\
corresponding to about 15 kpc. This is larger than the entire X-ray bright image observed
by \chandra\ (see below) hence all our \xmm\ related results refer to the total flux
of the source.

Our work makes use also of the {\it Chandra} data published by K03.
This observation was obtained on 2001
July 29 using the ACIS-S3 CCD (Obs. ID. 1590) and resulted with a
GTI of 37 ks.  We obtained the data from the {\it Chandra} archive
and reduced and analyzed them using CIAO3.2.1 (with CALDB3.0.0) and its standard
threads for extraction of extended and point like sources. We used
various spatial and energy filters to extract images and spectra as
detailed in the following sections. We note that 1$\arcsec$ corresponds to 0.49
 kpc at the source and the sizes of all the spatially resolved images discussed below
are  much larger than the instrument point spread function (PSF) which contains 80\% of the
power within 0.685\arcsec. Thus none of the linear sizes and volumes derived
below is affected by more than about 10\%; a small uncertainty considering the other unknowns
discussed below.

Since our observations were split into various segments that were separated by weeks to months, we
were able to study the time dependent of the source. We have carried out a standard analysis using
the high energy (less obscured) $E>6$ keV continuum. We found no variability to within about 10\%. We
note that Ptak et al. (2003) show the historical (1994--2002) light curve of the source
 where they detect variations by up to 40\%.

\section{Spectroscopic analysis}

\subsection{A comparison of the spectra of NGC\,6240 and M82}

M82 is a classical luminous starburst galaxy that has been studied by all X-ray satellites (Read \&
Stevens 2002 and references therein). The galaxy is known to have a prominent superwind most clearly
seen in $H_{\alpha}$ images as well as in X-ray lines (e.g. Stevens, Read and Bravo-Guerrero, 2003).
The \xmm\ based work of Read and Stevens and Stevens et al. claim a two region superwind with indications
for a hotter central region. They also suggested abundances that deviate from solar and some contribution to
the observed high energy continuum from a known point source (possibly a medium-size black hole)
 and perhaps several unresolved X-ray binaries.
The starburst origin of most of the observed X-ray emission is clearly indicated by the many
iron charge states and by the line ratios in several He-like ions (Read and Stevens 2002).

\begin{figure}
{\includegraphics[width=9.2cm]{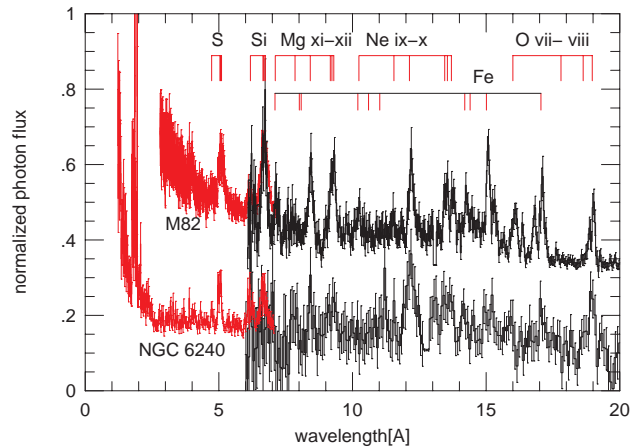}}
\caption
{Comparison of the \xmm\ spectra of M82 and NGC\,6240. The RGS spectra are shown in black and the pn
spectra in red. The flux scale of the two
has been arbitrarily shifted to allow easier comparison.
}
\label{RGS_1}
\end{figure}

Fig.~\ref{RGS_1} shows the RGS spectra of NGC\,6240 and M82. The M82 spectrum  was obtained from the
 \xmm\ archive and is based on the data presented in Read and Stevens
(2002) and in Stevens et al. (2003;  30 ks integration, 25 ks GTI). It was extracted
in the same way as the NGC\,6240 spectra.
The spectra shown here are normalized to allow easy comparison.
The similarity of the 5.5--20~\AA\ spectrum of the two galaxies is striking. All strong emission lines seen in
the spectrum of M82 are reproduced in the (noisier) spectrum of NGC\,6240 or are consistent with
suspected emission features. As explained above, the $\lambda > 20$~\AA\ RGS spectrum of NGC\,6240
cannot be used for obtaining line fluxes or even meaningful upper limits.

Line intensities for the two galaxies were measured from their RGS spectra by fitting simple Gaussians
to strong observed features. Theoretical wavelengths and zero intrinsic widths were used as constraints on the
measurements. The results are listed in Table 2 (note that the Read and Stevens 2002 and the Stevens
et al. 2003 papers do not list the line intensities used in their analysis of M82). Obviously, the uncertainties on
the line intensities in NGC\,6240 are larger due to the worse S/N of its RGS spectrum. An important issue
is the relative intensity of the various line components ($r$, $i$ and $f$) in He-like ions. These are very good
temperature indicators provided the effect of line pumping (sometimes referred to as continuum
fluorescence) is taken into account. Unfortunately, the RGS resolution is insufficient to provide
a clear separation
of the various He-like line components for ions with $Z>12$.
 Moreover, the S/N of our
NGC\,6240 RGS spectrum requires heavy binning thus we cannot trust individual line components. The
measurements of the \neix\ complex suggests it is dominated by the $r$ line, the signature of a high
temperature gas, but the uncertainty is large and we prefer using other types of analysis explained below.

\begin{deluxetable*}{lcccc}
\tablecolumns{5}
\tabletypesize{\footnotesize}
\tablewidth{330pt}
\tablecaption{Measured line fluxes \tablenotemark{a}
\label{line_flux}}
\tablehead{
\colhead{Line} &
\colhead{Wavelength$[\AA]$} &
\colhead{NGC\,6240} &
\colhead{M82} &
\colhead{Instrument\tablenotemark{d}} }
\startdata
\Kb\tablenotemark{b}     &1.75           &$15.5\pm1.1$           &-                        &pn+MOS \\
\feXXV\                  &1.85             &$81.4\pm7.9$           &$174.9_{-22.2}^{+29.9}$&pn+MOS/pn \\
\Ka\                     &1.94            &$112.1 \pm7.8$ &-                        &pn+MOS \\
\sXVI\tablenotemark{c}   &4.73            &$6.0 \pm-3.0$      &$59.2_{-14.3}^{+13.4}$ &pn+MOS/pn \\
\sXV\                    &5.04, 5.10     &$20.5 \pm2.8$&$240.8_{-30.7}^{+26.7}$  &pn+MOS/pn \\
\siXIV\                  &6.18            &$16.5 \pm2.1$&$126.0\pm63.5 $           &pn+MOS/RGS \\
\siXIII\                 &6.65            &$26.0 \pm1.9$&$519.0\pm150.0   $           &pn+MOS/RGS \\
\mgXII\                  &8.42            &$20.1\pm8.1 $          &$285.0\pm94.0  $           &RGS \\
\mgXI\                   &9.17            &$18.5\pm17.5 $          &$335.0\pm96.9 $           &RGS \\
\nex\                    &12.13           &$25.4\pm12.5 $         &$115.1\pm40.3 $          &RGS \\
Fe{\sc xxi}              &12.28           &$13.7\pm10.7 $         &$100.3\pm30.8 $          &RGS \\
\neix\                   &13.45           &$12.2 \pm 4.0  $       &$236.0\pm92.1 $           &RGS \\
Fe{\sc xx}, Fe{\sc xix}  &13.77, 13.79     &-                        &$65.2\pm23.3 $           &RGS \\
Fe{\sc xviii}            &14.21           &-                        &$63.7\pm16.4 $           &RGS \\
Fe{\sc xvii}, Fe{\sc xix}&15.01, 15.08   &$5.6\pm2.6  $          &$151.0\pm30.4 $           &RGS \\
Fe{\sc xvii}             &15.26           &$9.3\pm3.1  $          &$64.2\pm22.0  $           &RGS \\
Fe{\sc xviii}, Fe{\sc xix}&16.00, 16.07, 16.11&-                      &$63.4\pm21.3 $           &RGS \\
Fe{\sc xvii}              &16.78          &-                        &$41.4\pm15.0  $           &RGS \\
Fe{\sc xvii}              &17.05, 17.10  &$5.5\pm2.8 $          &$81.0\pm26.0   $           &RGS \\
\oviii\                   &18.97          &$12.1\pm3.6 $          &$69.9\pm27.3 $           &RGS \\
\enddata
\tablenotetext{a}{In units of $10^{-15}$ \ergcms.}
\tablenotetext{b}{Line intensity constrained to 0.114 the intensity of \Ka.}
\tablenotetext{c}{Flux and uncertainty estimated from integrating above a power-law fit to the local continuum.}
\tablenotetext{d}{Left is  for NGC\,6240 and right for M 82.}
\end{deluxetable*}
%

As evident from Table 2, the intensity ratios of most emission lines in the two spectra are very similar.
There is some tendency for the H-like/He-like  line
ratio of some elements to be somewhat larger in NGC\,6240. This is most noticeable
for the silicon lines. However, the silicon lines region is at the end of the RGS response and the
spectrum in this region
is particularly noisy. Thus the uncertainty on the flux of the \siXIV\ line is very large.
To improve the line measurement, and in order to add lines  outside
of the RGS range, we have included in the comparison the MOS and pn spectra of the two galaxies.
We fitted models to the data over the 1.7--10 keV range using XSPEC version 11.3.0.
The NGC\,6240 model consists of the following components:
a. An absorbed power-law.
b. A large column density line-of-sight neutral absorber.
c. An additional power-law which may be either the reflected high energy continuum
(in which case the two were forced to have the same slope) or due to local X-ray sources. The most likely
contributers in this case are X-ray binaries which are known to have harder X-ray continua. The power-law
in this case was forced to have a flatter slope. An additional possibility is a very high temperature thermal continuum mimicking
a power-law. As shown below, such temperatures are inconsistent with the temperatures derived from the
emission lines.
d. The following emission lines (all Gaussians with zero-widths which gives the instrument resolution when used in XSPEC):
1.865 keV (\siXIII), 2.006 keV (\siXIV), 2.461 keV (\sXV), 2.622 keV (\sXVI), 6.400 keV (Fe{\sc~ i-xvii}, hereafter \Ka),
 6.700 keV (\feXXV), 6.96 keV (\feXXVI) and 7.100 keV (Fe{\sc~ i-xvii}, hereafter \Kb).
All line energies were fixed at those rest energies and the \Ka/\Kb\ ratio was forced to its known theoretical value.
$\chi^2$ minimization is obtained by fitting, simultaneously, the pn and the two MOS spectra.

The results of the fitting process are as follows:
The two powerlaws with forced identical slopes gave $\Gamma=1.70 \pm 0.01$.
The two independent power-laws gave $\Gamma=1.76\pm0.02$ for the absorbed one and
 $\Gamma=1.19\pm0.02$ for the
 (``binary'') unabsorbed power-law.
The absorbing columns are $N_H=1.37 \pm0.02 \times 10^{24}$ \cmii\ for the first case and
$N_H=1.35 \pm0.03 \times 10^{24}$ \cmii\ for the second case. The slopes and the column densities
 are in rough agreement with the
Netzer et al. (1998), Vignati et al. (1999) and Ptak et al. (2003) earlier results.
In all cases, acceptable fits were obtained (with reduced $\chi^2$ in the range  0.79--0.92).
The observed EWs for the highest energy lines are 1.09 keV for the \Ka\ line and 0.81 keV for the \feXXV\ line.
 Both fits do not require any \feXXVI\ line although we cannot exclude a weak line
that contributes up to  30\% to the feature identified here as \Kb.

We have made no attempt to fit the entire 0.3--10 keV pn and MOS spectra of NGC\,6240
in a way similar to several earlier works. Such models are necessarily limited by the low spectral
resolution of CCD-type detectors that results in a degeneracy with respect to the composition and
temperature of the emitting gas. As explained in \S1, several earlier works using such
fits resulted in erroneous conclusions.
Our findings are based as much as possible on the RGS data at
long wavelengths, where such a degeneracy does not exist. The pn-MOS fit is therefore only at high
energies. Nevertheless, we believe that the \siXIII\ and  \siXIV\ line fluxes obtained
from such fits are more reliable than the ones obtained from the RGS spectrum, which is very
noisy at those energies. Therefore they are the ones listed in Table 2.

\begin{figure}
{\includegraphics[width=9.2cm]{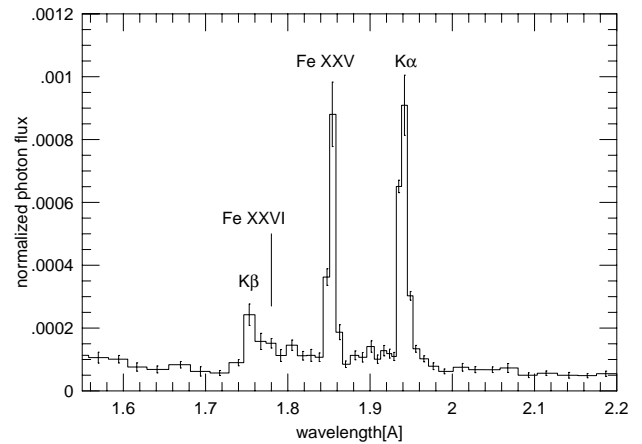}}
\caption
{The iron-K complex in NGC\,6240 (combined pn and MOS spectra). }
\label{pn_spectrum}
\end{figure}

A small part of the pn spectrum obtained with the tabulated line intensities
is added to Fig.~\ref{RGS_1} on the same scale as the RGS spectrum.
The iron-K region (unfolded spectrum, combined pn and MOS data with the line fluxes listed in Table 2)
is shown in Fig.~\ref{pn_spectrum}.

The fit to the pn spectrum of M82 was done for the purpose of measuring lines with wavelengths shorter than 6~\AA.
This was achieved by fitting zero-width lines to a continuum that was determined in two different ways:
 a. a single powerlaw.
b. the multi-temperature model of Read and Stevens (2002) with zero abundance for all heavy elements.
The results of the two methods are similar and the line intensities reported in Table 2 are
the ones obtained with the second continuum.  All other lines, with longer wavelengths, were measured by fitting
local continua and Gaussian-shaped lines to the RGS spectrum of the source as done for NGC\,6240.
We note that the purpose of our work is to model the spectrum of NGC\,6240,
hence the only aim of discussing the M82 spectrum is to provide a basis for comparison
for  NGC\,6240.

\subsection{Thermal gas in NGC\,6240}
The great similarity of the 4--20 \AA\ spectra of NGC\,6240 and M82 suggests that the large
scale X-ray emission in NGC\,6240 is dominated by a massive starburst.
Given the strong Ne{\sc ix} and Fe{\sc xvii} lines, that are typical of low temperature plasma, and the prominent
 silicon and sulphur H-like lines typical of much higher temperature gases,
we suggest that there are at least two regions with very different temperatures in this source.
Boller et al. (2003) reached a similar conclusion by comparing their \xmm\ observations with the spectrum
of NGC\,253, another well known starburst galaxy.
These findings are also
supported by the multi-energy maps of K03 that suggests that the gas emitting the 0.5--1.5 keV flux is mostly
in the outer parts of the system, further away than 2--3 kpc from the two nuclei. According to them,
the gas emitting the 1.5--2.5 keV radiation
is closer in and there is a small inner part that emits most of the $E>5$ keV radiation.

The parameters of the models presented below are the
temperatures, the abundances of oxygen, neon, magnesium, silicon, sulphur and iron, and the emission measure,
\begin{equation}
EM(T) = \int n_e n_H \epsilon dV \,\,\, \cc ,
\end{equation}
where $\epsilon$ is a volume filling factor.
We have calculated several low density thermal plasma models for a range of temperatures and
compositions and compared them with the line list given in Table 2.
  All models shown here, as well as the photoionization
models discussed in \S3.3, were computed with ION2004, the 2004 version of the code ION (Netzer et al. 2003
and references therein). The definition of ``solar composition'' as used in these models for the elements
under study here is
N(O)/N(H)=$5 \times 10^{-4}$,
N(Ne)/N(H)=$1 \times 10^{-4}$,
N(Mg)/N(H)=$3.5 \times 10^{-5}$,
N(Si)/N(H)=$3.5 \times 10^{-5}$,
N(S)/N(H)=$1.6 \times 10^{-5}$ and
N(Fe)/N(H)=$4 \times 10^{-5}$.
An important addition to the earlier version of the code is the inclusion of more
realistic collision strengths for various iron ions and the better treatment of the various processes
in Fe{\sc xvii} (Doron and Behar 2002).

The first stage of the analysis includes re-examination of the spatial distribution of the emitting gas.
This was done by separating images obtained from the
\chandra\ archive into narrow energy bands around strong lines and by extracting \chandra\ ACIS-S spectra.
We found that most of the flux in the \oviii\ line
is emitted in the outskirt of the
source, in a few (4--5) localized regions outside a radius of about 2.1 kpc. These ``lobes'' are clearly seen in several of the
K03 images. The \nex\ line originates, in almost equal amounts, in the lobes and in a spherically looking
central part with a radius of 2.1 kpc.
We identify one component of the thermal gas model with the lobes
and name it ``the outer zone''. The temperature of this gas is probably between the peak emissivity
temperatures of the \oviii\ and \nex\ lines ($3-8 \times  10^6$ K). We also identify a higher temperature,
spherical looking zone
showing some \nex\ emission and most of the \mgxii, \siXIV\ and \sXV\ line emission.
By combining all pixels showing \nex\ emission {\it outside} of the 2.1 kpc radius, we can evaluate the
 fraction of the volume emitting this line. Our two dimensional line emission maps translate to roughly 115 kpc$^3$ which is about
half the volume between 2.1 and 4 kpc.
The inner zone that shows emission of \nex\ and higher ionization species has a
volume of about 39 kpc$^3$.

  Second, we examined the intensity ratios of the strongest \feXVII\ lines relative
  to \oviii. The emissivities of these lines peak very close to the emissivity of the \nex\ lines,
 but unlike the \oviii\
  and \nex\ lines, it drops sharply above about $10^7$~K.
  We found that at  temperatures of order $10^7$~K, and with Fe/O solar to twice solar,
  a collisional model gives a satisfactory fit to the relative intensities of the \feXVII\
  and \oviii\ lines. Thus the thermal gas model of NGC\,6240 requires
  between solar and $2 \times$solar Fe/O. We did not attempt any model where different temperature
  components have different metalicity as the data do not warrant such an attempt.

\begin{figure}
{\includegraphics[width=9.2cm]{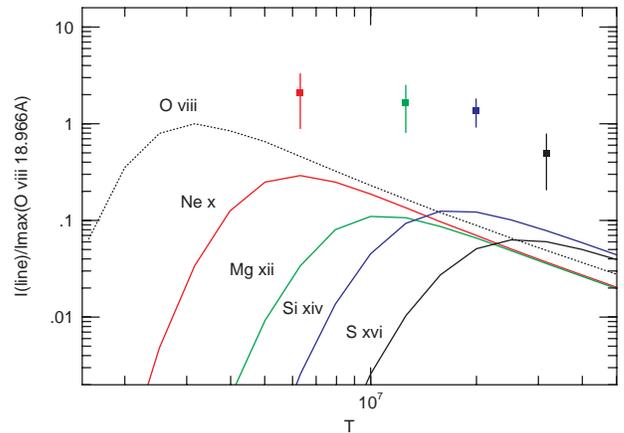}}
\caption
{Thermal line emission as a function of electron temperature for solar composition plasma.
The curves show computed $L_{\alpha}$ line
fluxes for various H-like ions relative to {\it the peak} emissivity of the \oviii\ $L_{\alpha}$ line.
Points with error bars give the observed ratios in NGC\,6240.
}
\label{thermal_lines}
\end{figure}

  Third, we investigated the higher temperature component which is required to explain
  the strong \siXIV, \sXVI\  and possibly \feXXV\ lines. The
  observed line ratios reflect the metalicity, temperature and EM of the gas in the various regions.
  This can be illustrated with the help of the generic 
  solar composition model shown in Fig.~\ref{thermal_lines}. The figure shows all the H-like $L_{\alpha}$
  lines
  observed by us relative to the peak emission of the \oviii\ $L_{\alpha}$ line
 as a function of temperature. 
This is the best way to illustrate the intensities of collisionally excited lines that are emitted at
temperatures close to their peak emissivity temperature. Obviously, the real situation in NGC\,6240 is
more complicated but the diagram helps illustrate the main point regarding the need for temperature distribution
in the source.

As evident from  Fig.~\ref{thermal_lines}, 
 the assumption that most line emission occurs at temperatures close to the peak emissivity
  temperature of the various ions clearly indicates that neon, magnesium, silicon and sulphur require 
either larger abundances
  and/or  much larger EM in their production regions compared with oxygen.
  This confirms the findings based on the imaging analysis that there must be 
 an inner central region with  temperatures that are large enough ($ T > 1.5   \times 10^7$ K)
to explain the intensities of the magnesium, silicon and sulphur
lines. If the temperature in some parts of this region exceeds $\sim 4 \times 10^7$~K, it can also explain the
observed \feXXV\ line.

As explained, the line emissivities shown in
Fig.~\ref{thermal_lines} suggest possible  over
abundances of Ne/O, Mg/O, Si/O and S/O. Before investigating
this in detail  we note that there are indications for high abundances in M82, our comparison source
(the abundances found by Read and Stevens (2002) for M82  are not entirely consistent with
the Stevens et al. (2003) results although the two investigations are based on the same \xmm\ observation).
This is also the conclusion of the recent Origlia et al. (2004) analysis of the source.
 We have no way to scale the metal abundances in NGC\,6240
relative to hydrogen because of the very large uncertainty on the intensity of the low temperature
thermal continuum. Thus, all the abundance analysis presented here is relative to oxygen.
Having no way to deduce the oxygen abundance, we assumed solar O/H. While none
 of our conclusions depend on this assumption, we note that it is
consistent with the Read and Stevens (2002) finding for O/H in M82. 

\begin{figure}
{\includegraphics[width=9.2cm]{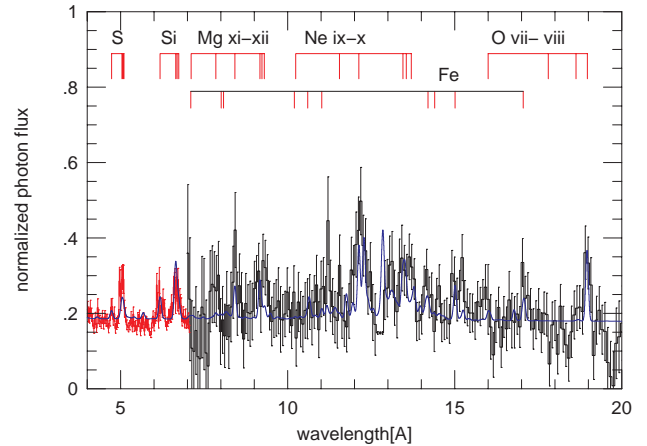}}
\caption
{A three-zone model fit to the spectrum of NGC\,6240. The black and red lines are the RGS and pn spectra shown  in
Fig. 1 and the blue line is the three-temperature model. The RGS data is binned to 0.08\AA\ and the model assumes
Gaussian emission lines with FWHM corresponding to to this width.}
\label{RGS_2}
\end{figure}

We carried out a combined abundance and temperature analysis aimed to find the combination of temperature,
EM and composition that best fits the spectrum of NGC\,6240. Given the large uncertainty on most line fluxes, we only
investigated three zone models and four different compositions ranging from solar to $2.5\times $solar
metalicity. We treat the elements in two groups, one containing Ne, Mg, Si and S, and one containing Fe.
 We have constructed a grid of thermal models covering the temperature range $10^6 - 10^8$~K and changed the
 assume EMs for the various zones over a large range. The results were
 tested against seven observed line ratios (seven line intensities relative to the \oviii\ 18.96\AA\ line):
 $L_{\alpha}$ of H-like neon, magnesium, silicon, and sulphur, the He-like \feXXV\ line,
 the \feXVII\ line at 15.0~\AA\ and the upper limit on \feXXVI.
 The results of the analysis are:
 \begin{enumerate}
 \item
 No solar abundance model can produce all the observed line ratios with their uncertainties. The best
 solar composition model can fit only 4 of the seven line ratios.
\item
Models with $1.5-2.5 \times$solar for Ne/O, Mg/O, Si/O and S/O can fit all line ratios to withing the uncertainties
provided Fe/O is between solar and $2.5 \times$solar.
The higher abundance cases (2 and $2.5 \times$solar) produce significantly better fits and a larger variety of
acceptable combinations of temperatures and EMs.
\item
The spatial distribution of the \oviii\ emission can be used as an additional constraint on the model.
The assumption 
that at least 60\% of the total line emission comes from the outer zone (as our analysis clearly indicates)
 does not change any of the
previous conclusions. Assuming this fraction is more than 85\%, which is also consistent with the available data,
forces the temperature of the outer region
to above $10^7$~K and gives poorer fits mostly for the \feXVII\ lines. 
Unfortunately, accurate determination of this fraction using the available data is difficult to obtain. 
For example, we cannot exclude the possibility that
some of the \oviii\ emission originates even further from the center than indicated by the \chandra\ images.
Such a component is likely to have temperatures much below $10^7$ K.
\item
There is no unique solution to the temperature-EM distribution in the source. In fact, the temperatures and EMs
in the three zones can change by up to 0.1 dex while still being inside the range allowed by the observations.
Thus, there are dozens of solutions within these uncertainties and no statistically sound way to tell them apart.
\end{enumerate}
The conclusion of the analysis is that a multi-temperature, multi-zone thermal plasma model can explain all observed emission
lines except for the \Ka\ and \Kb\ lines, to a reasonable accuracy.
Table~\ref{EM} shows two sets of EMs and temperatures that successfully reproduce all seven line ratios
 for a case of solar Fe/O and twice solar for the other elements, and a second case
of $2 \times$solar for Ne/O, Mg/O, Si/O, S/O and Fe/O.
Table~\ref{EM_lines} shows the calculated line ratios for the two cases and compares them with the observations.
In the rest of the paper we focus on the first case (model A in the table).
Fig.~\ref{RGS_2} shows a fit of this model to the 4--20~\AA\ spectrum of NGC\,6240.

\begin{deluxetable*}{lcccccc}
\tablecolumns{8}
\tabletypesize{\footnotesize}
\tablewidth{300pt}
\tablecaption{Temperatures (in K) and emission measures (in \cc) for  thermal models
\label{EM}}
\tablehead{
\colhead{Case} &
\colhead{$T_1$} &
\colhead{$EM_1$} &
\colhead{$T_2$} &
\colhead{$EM_2$} &
\colhead{$T_3$} &
\colhead{$EM_3$} }
\startdata
Model A& $10^7$  & $1.9 \times 10^{64}$& $1.6 \times 10^7$ & $1.4 \times 10^{64}$ & $6.3 \times 10^7$ & $5.4 \times 10^{64}$  \\
Model B & $ 8 \times 10^6$  & $1.2 \times 10^{64}$ & $1.6 \times 10^7$ &  $2.0 \times 10^{64}$ & $ 5 \times 10^7$ & $3.5 \times 10^{64}$  \\
\enddata
\end{deluxetable*}

\begin{deluxetable*}{ccccccc}
\tablecolumns{7}
\tabletypesize{\footnotesize}
\tablewidth{300pt}
\tablecaption{Line ratios relative to \oviii\ $\lambda 18.966$ for the models in Table~\ref{EM}
\label{EM_lines}}
\tablehead{
\colhead{} &
\colhead{\oviii} &
\colhead{\nex} &
\colhead{\mgXII} &
\colhead{\siXIV} &
\colhead{\sXVI} &
\colhead{\feXXV} }
\startdata
Observed& 1.0 & $2.10 \pm 1.21$ & $1.66 \pm 0.83$ & $1.36 \pm 0.44$ & $0.50 \pm 0.30$ & $6.73 \pm 2.10$ \\
Model A & 1.0 & $1.60   $ & $1.14 $ & $1.22 $  & $0.56 $  & $5.99 $\tablenotemark{a}   \\
Model B & 1.0 & $1.55 $   & $0.93 $ & $ 1.16 $ & $ 0.53 $ & $6.26 $\tablenotemark{b}  \\
\enddata
\tablenotetext{a}{Solar Fe/O and twice solar for Ne/O, Mg/O, Si/O and S/O.}
\tablenotetext{a}{Twice solar for Ne/O, Mg/O, Si/O, S/O and Fe/O.}
\end{deluxetable*}

There are two fundamental uncertainties related to the suggestion that the 2.1 kpc radius region
can be separated into two sub-zones with 
the temperatures found earlier. They are both
related to the role of photoionization close to the two active nuclei. The
 first is the origin of the \feXXV\,1.85~\AA\ (6.7 keV) line. This line
can be produced by hot thermal plasma, as shown in our calculation,
 or by recombination in a colder photoionized gas, as seen in several
well studied type-II AGN.
 The latter case is investigated in \S3.3.
The second uncertainty  is related to the relatively strong 2--5 keV power-law continuum
that is required by the spectral fit. As explained in \S3.1, such a continuum can be produced by a very high
temperature gas, by a local population of X-ray binaries
or by reflection of the central power-law continuum.
The second possibility has been discussed in several recent papers, see e.g. Persic and Rephaeli (2002).
The predicted continuum in this case is very hard, with $\Gamma \sim 1.2$.
The third mechanism is known to dominate
the soft X-ray emission of other type-II AGN at energies where the central source is obscured. In such cases,
the reflected continuum  is seen from an extremely small, usually unresolved part of the source.

\begin{figure}
{\includegraphics[width=9.2cm]{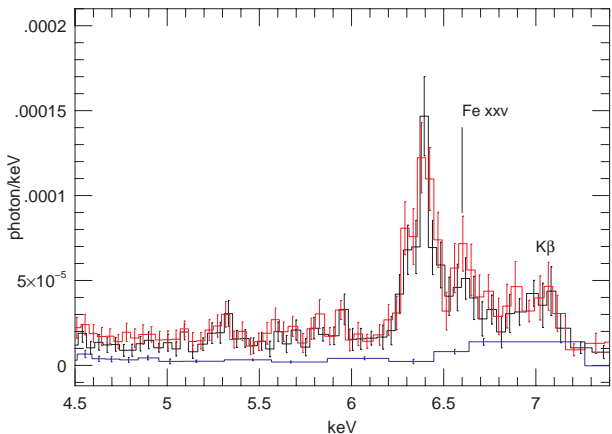}}
\caption
{\chandra\ spectra of the outer (blue), central (red, all emission within 2.1 kpc from the center)
and innermost (black) zones in NGC\,6240. Note that
most of the 4--6 keV continuum comes from the innermost part while
the \feXXV\ line seems to originate mostly in the central region.
}
\label{chandra_three_zones}
\end{figure}

To study this issue we have  extracted the \chandra\ spectra of three separate regions of the source:
the outer zone between 2.1 and 4 kpc, the central zone with a radius of 2.1 kpc,  and an innermost zone encompassing
the two nuclei. The latter is seen as an elliptical region and 
can be described by an ellipsoid with a major axis of
 1.4 kpc and two other axes of 0.7 kpc each. The larger dimension is along the line connecting the two nuclei
and the separation between the nuclei is roughly the size of the semi-major axis.
The corresponding volume is about 0.36 kpc$^3$. 
The high energy parts of these spectra are shown in
Fig.~\ref{chandra_three_zones}.

The comparison of the three spectra confirms the general suggestion that the outer
zone contributes most to the low energy emission while the inner zones produce spectra
that are typical of a much higher temperature gas.
Unfortunately, some features in the \chandra\ spectra are
different from those seen in the combined pn+MOS spectra for reasons that we do not understand.
This is seen mostly as excess emission between about 2.1 and 2.4 keV. The problem  has already been noted by Ptak et al. (2003)
in their modeling of the K03 data and was suspected to be a calibration error due
 to a well known mirror feature near the gold edge. We are not
 sure how to confirm this idea since we used the best available calibration provided by the CXC (see \S2).
Given this uncertainty, we decided
 not to use the \chandra\ spectra for measuring line intensities and only to address, qualitatively, the
issues of the \feXXV\ line and the reflected continuum.

As seen in the full \chandra\ spectra (not shown here), and as
hinted in Fig.~\ref{chandra_three_zones}, there is a
 tendency for the relative intensity of the 2--5 keV continuum to increase inward, supporting the idea of
reflection in the photoionization region.
The opposite is true for the \feXXV\ whose flux seems to originate in an {\it intermediate
part} outside the elliptical photoionization region.
These findings are  marginal because of the difficulty in separating the two small zones in the center and
because of the problematic \chandra\ spectra. They are shown here in conjunction with the theoretical argument
presented below (\S3.3) and seem to support it.
The situation regarding the \Ka\ line is clearer and the diagram shows that the line originates in the inner zone.
The small elliptical zone is termed below ``the photoionized region''.

Given the qualitative impression based on the \chandra\ imaging and Fig.~\ref{chandra_three_zones},
and the detailed modeling of the thermal
gas, we  suggest the following three zone structure for the starburst dominated regions in NGC\,6240.
The first part is the one addressed earlier, outside a 2.1 kpc radius.
The electron temperature there is $8-10 \times 10^6$ K and the volume occupied by the line emitting gas is about 115 kpc$^3$.
Inside the 2.1 kpc radius there is an inner spherical part (hereafter ``the inner zone'')
where the temperature is $5-6 \times 10^7$ K and  where most of the \feXXV\ is emitted. Outside of this
zone there is an outer
shell (hereafter the ``intermediate zone'') where the temperature is $\sim 2 \times 10^7$ K and where
most of the magnesium and silicon line emission takes place.
We also suggest that the small-volume innermost photoionized zone is part of the inner thermal plasma zone.
We adopt, somewhat arbitrarily, a radius that divides the 2.1 kpc sphere into two concentric regions
of equal volume (19.5 kpc$^3$). This division is not based on emission line
imaging and is, therefore, rather  uncertain.
However, the general abundances derived here, and the total hot gas mass deduced below,
are rather insensitive to the exact division.

\subsection{Photoionized gas in NGC\,6240}

The two strong point-like  X-ray continuum sources observed in NGC\,6240 can ionize low density gas many 
pc away from the center. Such gas will
produce fluorescence and recombination iron lines.
The intensity of the lines depends on the covering factor and the column density of the illuminated gas and
 its level of ionization.
The strong 6.4 keV \Ka\ iron line (Fe{\sc~i-xvii}) observed in NGC\,6240 cannot be produced by the
starburst gas and must be due to fluorescence. Indeed, the K03 findings and our own work
 (Fig.\ref{chandra_three_zones})
are consistent with the production of this line only in the innermost photoionized  part of the source.
In contrast, the \feXXV\ 1.85~\AA\ line can be produced either by hot gas in starburst regions or in high ionization
photoionized gas. A good example of the later is the case of
NGC\,1068 (Marshall et al., 1993; Netzer \& Turner 1997; Kinkhabwala \etal, 2002).
Below we discuss this possibility for the special case of NGC\,6240.

\begin{figure}
{\includegraphics[width=9.2cm]{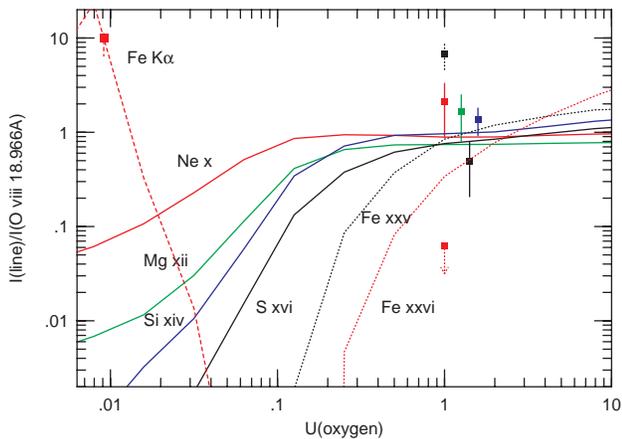}}
\caption
{Line ratios relative to the  \oviii\ $L_{\alpha}$ line for photoionized gas with twice solar abundance
for Ne/O, Mg/O, Si/O and S/O and solar O/H and Fe/O.
 The column density is $10^{22}$ \cmii\ except for the Fe-\Ka\ line where the column density is $10^{23}$ \cmii.
Points with error bars, or upper limits,
 represent the data in Table 2 and are arbitrarily located along the horizontal axis to
illustrate the difficulty to obtain the observed line ratios for a specific value of $U(oxygen)$.}
\label{ph_model}
\end{figure}

Fig.~\ref{ph_model} shows the results of photoionization calculations that predict the intensities
of  the \nex, \mgxii, \siXIV, \sXVI, \Ka, \feXXV\ and \feXXVI\ lines  relative to the intensity of the \oviii\ $L_{\alpha}$ line.
The intensity ratios are shown as a function
of the ionization parameter $U(oxygen)$ which is the photon to the hydrogen density ratio
defined over the energy range 0.54--10 keV (e.g. Netzer et al. 1998). The calculations were done
for the composition found here (model A), a power-law continuum with $\Gamma(0.1-50\,\, {\rm keV})=1.8 $,   a
 small UV bump (the exact shape of which is not relevant to the present calculations), and
a range of column densities between $10^{20}$ and $10^{23}$ \cmii. The diagram shows the $10^{22}$ \cmii\
case but the main conclusions are the same for all.

The new calculations show that many line ratios involving H-like ions are in rough agreement with the observations
of NGC\,6240 given enough freedom in the choice of $U(oxygen)$. This is not surprising since the most
important temperature diagnostics for the elements considered here, the $r$ and $f$ lines of the He-like ions,
 are missing from the diagram. The $f/r$ ratio can, by itself, separate thermal from photoionized gas.
As explained earlier,
the combination of poor resolution and S/N prevented us from using such ratios in a meaningful way thus we resort
to alternative methods. The next important conclusion is that the calculated \feXXV\ line (the sum of all line components)
 is too weak relative to its observed value even if we allow for a higher iron abundance.
Furthermore, the calculations show
that photoionized gas producing strong \feXXV\ must also produce
  strong \feXXVI\  which is in contradiction with the observations (see Fig.~\ref{ph_model}).
  Thus,  the photoionization origin of the \feXXV\ line in NGC\,6240
is practically ruled out by the calculations.

We also considered the  production of \Ka\ and \Kb\ lines. These
were computed in a similar fashion and the \Ka\ line is shown in Fig.~\ref{ph_model} except that
in this case we give its relative intensity for a column density of $10^{23}$ \cmii\ (the smaller
columns cannot produce the observed line intensity). Obviously, the line is produced in a much
lower ionization gas which is entirely consistent in terms of flux and EW with dozens of
type-II observations and with theoretical calculations (Netzer et al. 1998 and references therein).
Thus, the relative intensity of the lowest and the highest ionization iron lines, plus
the success of the spatially resolved thermal gas model, all suggest that in NGC\,6240,
\Ka\ and \Kb\  are the only strong emission lines produced in the photoionized gas.

\section{Discussion}
The new data and modeling of NGC\,6240 presented in this paper show an unusual ULIRG which is
perhaps the clearest case, so far, of a very luminous type-II AGN combined with a very high luminosity
X-ray starburst. The outer parts of the source, beyond about 2.1 kpc, are clearly dominated by the starburst
which is very similar in its properties to the starburst observed in M82, except for the much  higher luminosity.
The innermost part, very close to the two active nuclei
(an ellipsoid of about $0.7 \times 0.7 \times 1.4$ kpc),
shows  clear signatures of photoionized gas;  a strong \Ka\ 6.4 keV line and possibly a reflected continuum.
 The region between the photoionized part and a radius of about 2.1 kpc
is dominated by a high temperature starburst with strong emission lines of H-like and He-like magnesium,
silicon and sulphur and He-like lines of \feXXV.
The analysis suggests a temperature gradient towards the center which
we model as a three zone starburst region.

NGC\,6240 is among the brightest known X-ray luminous starbursts. The integrated 0.5--5 keV luminosity,
plus the luminosity of the \feXXV\ line, is about
 $1.3 \times 10^{42}$ \ergs. This is about a factor 50
 more luminous than the starburst in M82.  There are several other ULIRGs with comparable soft X-ray
 luminosities. For example, Franceschini et al. (2003) list two ULIRGs with 0.5--2 keV luminosity exceeding
$2 \times 10^{42}$ \ergsec. However, according to the same authors, most of the X-ray luminosity in those objects
is due to a power-law source which is likely to be  X-ray emission from numerous
binaries. The thermal gas emission in those cases is less than we find for NGC\,6240 (see their Table 3).
It is obvious that high resolution X-ray spectroscopy is key to the understanding of such sources and, so far, our
data set for NGC\,6240 is the only one that
allows clear separation of the AGN and the starburst components.

The emission measures found here (Table 3), combined with the volumes and temperatures of the
three thermal-gas zones, can be converted to density and mass of the ionized gas within the uncertainties on the filling factor.
The latter is a problematic issue since there are no direct measurements of the filling factor for this
or other starburst galaxies.
 Hydrodynamical simulations by Strickland and Stevens (2000) suggest that for X-ray bright galactic winds,
  $10^{-3} \leq \epsilon \leq 10^{-1}$. However, larger
 values are not excluded by any observation known to us. In the following we assume $\epsilon \sim 0.1$ but keep
 in mind the large uncertainty on its value.

Using the model A parameters from Table~\ref{EM} we get
$n_H \epsilon^{1/2} \simeq 0.07 $~\cc\ for the outer zone,
$n_H \epsilon^{1/2} \simeq 0.14$~\cc\ for the intermediate zone and
$n_H \epsilon^{1/2} \simeq 0.28 $~\cc\ for the innermost zone.
Assuming $\epsilon=0.1$ we get densities in the range 0.2--1 \cc, similar to those estimated by
Strickland and Stevens (2000) for the warm component in several of their  wind models.
The mass of the hot gas is given by
\begin{equation}
  M_{hot} \propto EM n_H^{-1} \propto \epsilon^{1/2} EM^{1/2} V^{1/2} \, .
\end{equation}
This gives for model A
$M_{hot} \simeq 1.9 \times 10^8 \epsilon^{1/2}$ \msun\ in the outer zone,
$M_{hot} \simeq 6.8 \times 10^7 \epsilon^{1/2}$ \msun\ in the intermediate zone and
$M_{hot} \simeq 1.3 \times 10^8 \epsilon^{1/2}$ \msun\ in the inner zone.
Under these assumptions, and for a uniform value of $\epsilon =0.1$ in all zones,
the total mass of thermal gas is about $1.2 \times 10^8$ \msun. The total mass derived assuming
 model B  is about a factor of 1.15 smaller.

The mass of hot gas obtained here can be compared with the expected properties of
a young system (e.g. colliding galaxies) undergoing violent star formation.
To estimate those properties we examined a starburst model computed by the code STARS
(A. Sternberg, private communication). The model
assumes a Kroupa (2001) IMF and an exponentially decaying star formation rate (SFR) with a characteristic time
of 20 mega-years (Myr).  The SFR is adjusted
to give the bolometric luminosity of NGC\,6240
 at its  peak which occurs after about 10 Myr. However, the peak is very broad and
the luminosity is larger than 80\% of the peak luminosity at all times between about 4 and 20 Myr.
The mass of cold  gas converted to stars in the first 10 Myr is about 2.5$\times 10^9$ \msun\
and the gas ejected by type-II supernovae is about 3$\times 10^7$ \msun\ after 6 Myr,
about $ 10^8$ \msun\ after 10 Myr and about $3 \times 10^8$ \Msun\ after 20 Myr.
All properties predicted by the theoretical model for the time interval of 10--20 Myr are in general agreement
with the properties deduced by our thermal plasma analysis assuming $\epsilon =0.1$.

The comparison of the observed and predicted hot gas mass is only valid if the ejected gas stays in the system
for the entire duration of the burst. This may not be the case for the system under study. Assuming a mean outflow
velocity for the thermal gas of $v_{out}$ \kms\ and a typical radius of 4 kpc,
we find a dynamical time of about
 $ (4000 {\rm ~ \kms} /v_{out})$ Myr, i.e. of the same order as the duration of the burst.
 NGC\,6240 is known to have a massive superwind associated with the central
starburst with a measured  $H_{\alpha}$ outflow velocity of about
1000 \kms\ (Heckman, Armus and Miley, 1990). This may indicate similar velocities for the X-ray gas in the source.
In such a case, our deduced hot-gas mass is probably a lower limit to
the total ejecta by the starburst provided most of the hot gas mass is carried outward by the wind
 (a rather uncertain assumption, see Strickland and Stevens, 2000).
 While it is not our intention to fully model
this complex scenario, we note that the uncertainty on the filling factor is large enough to
accommodate this case.
Incidentally, the dynamical timescale of the two nuclei in NGC\,6240
is of the same order, about 7 Myr (e.g. Tecza et al. 2000).

The results of the abundance analysis are more difficult to interpret. Our model suggests
super-solar metalicity relative to oxygen. This is problematic for $\alpha$-process elements and may indicate
that the model is over-simplified. The global properties are, however, in general agreement with
the simple starburst model we have used. In particular,
given type-II supernovae explosions as the main source of enrichment,  the composition of the ejected gas
 after about 10 Myr must be solar to within a factor of 2 for all heavy elements, including
iron. This is consistent  with our general finding.

Tecza et al. (2000) published the results of a detailed IR imaging spectroscopy of NGC\,6240. They found that
the K-band light is dominated by red supergiants formed in the two nuclei in starbursts triggered about 20 Myr ago.
They further deduced, based on their Brackett$-\gamma$ observation, a very short duration burst ($\simeq 5 \times 10^6$ ys)
which is already decaying. Despite the somewhat different starburst parameters of their model,
the general results are in rough agreement with our X-ray analysis regarding the age of the burst
and the mass of the ejecta. The finding of a very powerful starburst
also agrees with the ISO spectroscopy of Lutz et al. (2003).
Pasquali, Gallagher and de Grijs (2004) have modeled HST observations
of NGC\,6240 and constructed starburst models
for the various parts of the source. Their conclusions regarding the age and the star-forming rates are similar
to those found by us.
To conclude, our findings are consistent with a massive star forming event with an age of 10--20 Myr.
It remains to be seen whether other X-ray bright ULIRGs can give similar indications or whether NGC\,6240 is
an unusual member of the group.

Finally we note two general uncertainties in this and similar sources. The separation of the source into
a highly obscured AGN and completely exposed starburst regions is, of course,  simplified. Some of the innermost
highest temperature starburst regions can also be  obscured which will affect the observed line intensities.
We avoided this issue in the present study since using a global absorption correction factor is
meaningless and a proper analysis requires much better mapping of individual
starburst regions. This requires deeper \chandra\ imaging of NGC\,6240.

The second uncertainty is also related to the unknown geometry of the starburst gas. Our modeling assumes that the
hot thermal gas {\it is not} exposed to the central nonthermal radiation.
 This can be justified by the large line-of-sight obscuring
column. However, some geometries would allow the starburst gas to see this radiation which can
affect the level of ionization and the emitted flux in the starburst gas.
In particular, at E$>5$ keV, the cold gas may be transparent in several directions and thus the observed
intensity of the \feXXV\ line can be affected by continuum pumping which was not included in our modeling
of the highest temperature gas (but was included in the photoionization calculations).
More complex models that include all those processes must await better \chandra\ and \xmm\
observations of the source.

\acknowledgements
This work is supported by the Israel Science Foundation grant 232/03 and by the Bi-national Science Foundation
grant no. 2002-111.
We are grateful to Amiel Sternberg for allowing us to use his starburst code STARS and for useful discussions.
E. Behar has supplied useful atomic data. Many comments by an anonymous referee helped to improve the presentation
of this paper.
S.K.  acknowledges the financial support of the Zeff fellowship.


\begin{thebibliography}{}

\bibitem[Bogdanovi{\' c} et al.(2003)]{2003AJ....126.2299B} Bogdanovi{\'
c}, T., Ge, J., Max, C.~E., \& Raschke, L.~M.\ 2003, \aj, 126, 2299
\bibitem[Boller et al.(2003)]{2003A&A...411...63B} Boller, T., Keil, R.,
Hasinger, G., Costantini, E., Fujimoto, R., Anabuki, N., Lehmann, I., \&
Gallo, L.\ 2003, \aap, 411, 63
\bibitem[Dahlem et al.(1998)]{1998ApJS..118..401D} Dahlem, M., Weaver,
K.~A., \& Heckman, T.~M.\ 1998, \apjs, 118, 401
\bibitem[Dickey \& Lockman(1990)]{1990ARA&A..28..215D} Dickey, J.~M., \& 
Lockman, F.~J.\ 1990, \araa, 28, 215
\bibitem[]{}Doron, R., \& Behar, E., 2002, ApJ, 574, 518
\bibitem[Franceschini et al.(2003)]{2003MNRAS.343.1181F} Franceschini, A.,
et al.\ 2003, \mnras, 343, 1181
\bibitem[Genzel et al.(1998)]{1998ApJ...498..579G} Genzel, R., et al.\
1998, \apj, 498, 579
\bibitem[]{}
Heckman, T., Armus, L., \& Miley, G., 1990, ApJS, 78, 833
\bibitem[]{}
Kinkhabwala, A., \etal, 2002 ApJ 575, 732
\bibitem[Komossa et al.(2003)]{2003ApJ...582L..15K} Komossa, S., Burwitz,
V., Hasinger, G., Predehl, P., Kaastra, J.~S., \& Ikebe, Y.\ 2003, \apjl,
582, L15 (K03)
\bibitem
 [Kroupa(2001)]{2001MNRAS.322..231K} Kroupa, P.\ 2001, \mnras, 322, 231
\bibitem[]{}Levenson, N.A., Weaver, K.A., \& Heckman, T.M., 2001a, ApJS, 133, 269
\bibitem[]{}---------. 2001b, \apj, 550, 230
\bibitem[Lutz et al.(2003)]{2003A&A...409..867L} Lutz, D., Sturm, E.,
Genzel, R., Spoon, H.~W.~W., Moorwood, A.~F.~M., Netzer, H., \& Sternberg,
A.\ 2003, \aap, 409, 867
\bibitem[Marshall et al.(1993)]{1993ApJ...405..168M} Marshall, F.~E., et
al.\ 1993, \apj, 405, 168
\bibitem[]{}
 Max, C.E., Canalizo, G,  Macintosh, B.A.,  Raschke, L.,  Whysong, D.,  Antonucci, R.,
   \&  Schneider, G., 2004, (ApJ in press, astro-ph/0411590)
\bibitem[Netzer \& Turner(1997)]{1997ApJ...488..694N} Netzer, H., \& 
Turner, T.~J.\ 1997, \apj, 488, 694 
\bibitem[Netzer et al.(1998)]{1998ApJ...504..680N} Netzer, H., Turner,
T.~J., \& George, I.~M.\ 1998, \apj, 504, 680 
\bibitem[Netzer et al.(2003)]{2003ApJ...599..933N} Netzer, H., et al.\ 
2003, \apj, 599, 933
\bibitem[Origlia et al.(2004)]{2004ApJ...606..862O} Origlia, L., Ranalli,
P., Comastri, A., \& Maiolino, R.\ 2004, \apj, 606, 862
\bibitem[Pasquali et al.(2004)]{2004A&A...415..103P} Pasquali, A., 
Gallagher, J.~S., \& de Grijs, R.\ 2004, \aap, 415, 103
\bibitem[Persic \& Rephaeli(2002)]{2002A&A...382..843P} Persic, M., \&
Rephaeli, Y.\ 2002, \aap, 382, 843
\bibitem[Ptak et al.(2003)]{2003ApJ...592..782P} Ptak, A., Heckman, T.,
Levenson, N.~A., Weaver, K., \& Strickland, D.\ 2003, \apj, 592, 782
\bibitem[]{}
Read, A.M., Stevens, I.R., 2002, MNRAS, 335, L36
\bibitem[Sanders et al.(1988)]{1988ApJ...325...74S} Sanders, D.~B., Soifer, 
B.~T., Elias, J.~H., Madore, B.~F., Matthews, K., Neugebauer, G., \& 
Scoville, N.~Z.\ 1988, \apj, 325, 74
\bibitem[]{}
Sanders D., \& Mirabel 1996, ARAA, 34, 749 
\bibitem[Stevens et al.(2003)]{2003MNRAS.343L..47S} Stevens, I.~R., Read, 
A.~M., \& Bravo-Guerrero, J.\ 2003, \mnras, 343, L47
\bibitem[Strickland \& Stevens(2000)]{2000MNRAS.314..511S} Strickland,
D.~K., \& Stevens, I.~R.\ 2000, \mnras, 314, 511
\bibitem[Tacconi et al.(1999)]{1999ApJ...524..732T} Tacconi, L.~J., Genzel, 
R., Tecza, M., Gallimore, J.~F., Downes, D., \& Scoville, N.~Z.\ 1999, 
\apj, 524, 732
\bibitem
  [Tecza et al.(2000)]{2000ApJ...537..178T} Tecza, M., Genzel, R., 
Tacconi, L.~J., Anders, S., Tacconi-Garman, L.~E., \& Thatte, N.\ 2000, 
\apj, 537, 178
\bibitem[Turner et al.(1997)]{1997ApJS..113...23T} Turner, T.~J., George, 
I.~M., Nandra, K., \& Mushotzky, R.~F.\ 1997, \apjs, 113, 23 
\bibitem[Vignati et al.(1999)]{1999A&A...349L..57V} Vignati, P., et al.\ 
1999, \aap, 349, L57 


\end{thebibliography}
\end{document}